# Spectrum Sensing in Low SNR Regime via Stochastic Resonance

Kun Zheng, Husheng Li, Seddik M. Djouadi and Jun Wang

***Abstract*—Spectrum sensing is essential in cognitive radio to enable dynamic spectrum access. In many scenarios, primary user signal must be detected reliably in low signal-to-noise ratio (SNR) regime under required sensing time. We propose to use stochastic resonance, a nonlinear filter having certain resonance frequency, to detect primary users when the SNR is very low. Both block and sequential detection schemes are studied. Simulation results show that, under the required false alarm rate, both detection probability and average detection delay can be substantially improved. A few implementation issues are also discussed.**

## I. Introduction

In cognitive systems [1], spectrum sensing is one of the key issues. In order to effectively use the scarce spectrum resource, *secondary users* (SUs), which have no license to the frequency spectrum band, need to opportunistically share the spectrum with *primary users* (PUs) having license. An important requirement in cognitive radio systems is that SUs should not cause significant interference to PUs. This requires that SUs have the capability to sense the spectrum environment periodically and detect PUs' presence reliably.

The most challenging problem in spectrum sensing is how to detect very weak signals of PUs, or in other words, how to improve the signal-to-noise ratio (SNR) of received signal. One of the reasons why it is necessary to sense weak signals in cognitive radio systems is the existence of the *Hidden Primary User Problem* [2]. An example is illustrated in Fig. 1, where the PU receiver suffers substantial interference from the SU transmitter while the SU transmitter is unaware of the existence of the PU receiver. In traditional wireless networks, the hidden user problem can be alleviated by a clear-to-send (CTS) signaling from the receiver. However, in cognitive radio systems, PU systems cannot be modified and the risk of violating a dummy PU receiver cannot be alleviated via signaling. The only reliable approach is to improve the sensitivity of the SU's spectrum sensing such that it can detect the PU transmitter's signal once it is located within the interference range of the PU receiver. Thus it requires SUs to detect weak signals of PUs under required sensing time and achieve desired probabilities of detection and false alarm. For example, in 802.22 standard, the signal of PU must be detected in no more than 2 seconds with less than $10\%$ probability of false alarm and greater than $90\%$ probability of detection. Meanwhile, the IDT (Incumbent Detection Threshold) are -107 dBm and -116 dBm for wireless microphones and TV services respectively [3].

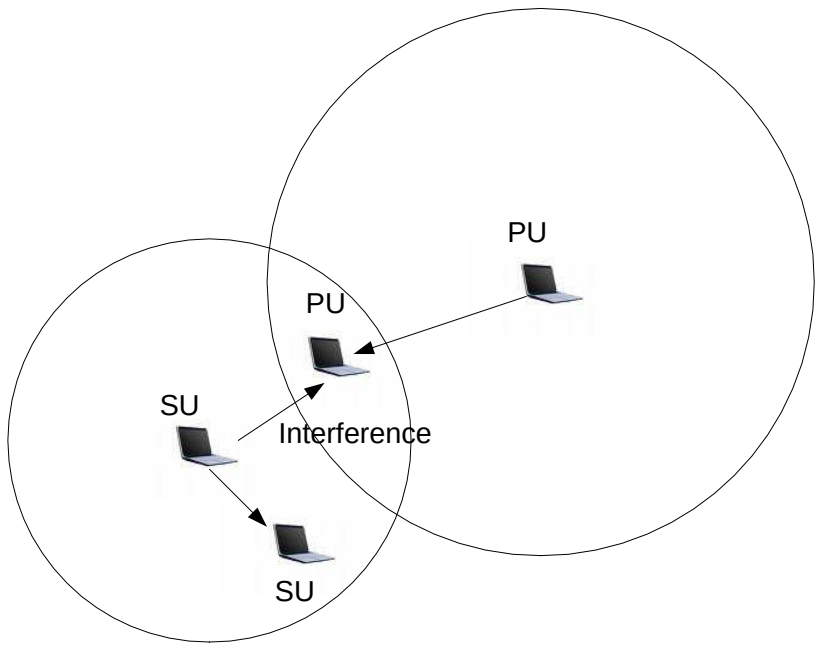

Fig. 1. Illustration of Hidden Primary User Problem

Several typical approaches can be used for spectrum sensing in cognitive radio systems:

- Energy Detection: It is extensively used in radiometry and is the most popular approach of spectrum sensing due to its low computational and design complexities [4] [5] [6]. Although energy detection based spectrum sensing is simple to implement and can be applied to any type of PU signal, there are still many drawbacks: (1) The energy detector does not distinguish signals between PUs and other SUs. Thus it may result in false alarm due to interference from other SUs' transmission, especially when SUs are not well time synchronized. (2) The usage of energy detection is limited by $SNR_{wall}$ [7], which results from the *noise uncertainty*. Under this $SNR$ threshold, the signal is found to be completely non-detectable by energy detection [7].
- Feature Detection: If we already know the specific characteristics of primary signal, such as cyclostationarity, pilot, radio identification and etc., feature detection can be applied [5] [8] [9]. For example, ATSC signal has many features that can be applied to feature detection. The main problem for feature detection method is the required large number of samples, thus incurring significant detection delay.
- Matched Filter: When transmitted signal is known at the

K. Zheng, H. Li and S. M. Djouadi are with the department of electrical engineering and computer science, the University of Tennessee, Knoxville, TN. J. Wang is with the National Key Lab of Communications, University of Electronic Science and Technology, Chengdu, P. R. China. This work was supported by the National Science Foundation under grant CCF-0830451, High-Tech Research and Development Program of China under grant 2007AA01Z209 and National Basic Research Program (973) under grant 2009CB320405.

TABLE I
COMPARISON OF TRADITIONAL DETECTION APPROACHES AND WHY SR CAN BE APPLIED

| | Energy Detection |
|---|---|
| Pros | simple to implement, applicable to any signal type |
| Cons | suffer to low SNR |
| | can't distinguish signal and interference |
| How SR works | increase input SNR to overcome $SNR_{wall}$ |
| | Feature Detection |
| Pros | capture specific characteristics of PUs signal |
| Cons | may incur significant detection delay |
| How SR works | may help to reduce detection delay |
| | Matched Filter |
| Pros | short time to achieve high processing gain |
| Cons | need perfect knowledge of the PUs signal |
| How SR works | require less information of PU signal |

receiver, the optimal linear detector is a matched filter [10]. It has been shown that a matched filter requires $O(1/SNR)$ samples to achieve a desired probability of detection error constraint [7]. On the other hand, if any feature of PU signal is unknown or mismatched, such as bandwidth, frequency, modulation type and order, the matched filter detector will be substantially degraded.

In this paper, we propose to use *stochastic resonance* (SR) filter [11] to combat the challenge of spectrum sensing, based on the assumption of periodicity of PU signal. SR was proposed by Benzi and his co-workers to explain the Earth's climate change in 1981. Essentially, SR is a nonlinear physical phenomenon. Due to the nonlinearity, a SR filter has a specific resonance frequency. When the input signal's frequency matches that of the SR filter, resonance happens and the signal on this frequency point will be significantly enhanced, thus improving the SNR. Note that the periodicity can be found in typical cognitive radio systems, e.g. the pilot in DTV systems or the periodicity of autocorrelation function in cyclostationary signals.

In contrast to the energy detection, the SR filter utilizes the *a priori* information about periodicity, thus substantially outperforming the energy detection, especially in the low SNR regime (as will be seen in the numerical simulations). Compared with the feature detection and matched filter, the SR has the much less requirement on the *a priori* information of PU system. Thus, SR filter is a good candidate for spectrum sensing in low SNR regime. The main advantages and disadvantages of traditional approaches (energy detection, feature detection and matched filter) are summarized in Table I, in which we explained how the SR alleviates the disadvantages of the traditional approaches.

The remainder of the paper is organized as follows. The system model is briefly reviewed in Section II. Stochastic Resonance is introduced in Section III. The proposed SR filter and the application in spectrum sensing are discussed in Section IV. Numerical results and conclusions are provided in Sections V and VI, respectively.

## II. System Model

In spectrum sensing, the detection problem is equivalent to distinguishing the following two hypotheses:

$$\begin{aligned} H_0 &: Y(n) = W(n) \\ H_1 &: Y(n) = S(n) + W(n), \end{aligned} \tag{1}$$

where $n$ is the observation index. $W(n)$ is the additive white Gaussian noise (AWGN). $S(n)$ is the PU signal to be detected. For simplicity, we assume that $S(n)$ is a sinusoidal signal, which is justified by the sinusoidal pilot in DTV system. For general case, we can always find periodic signal in PU systems. $H_0$ means that there is no PU in the channel and SUs are allowed to use the channel. $H_1$ means that PUs are present and SUs should quit from the channel. For block detection (the number of observations is fixed), the performance is measured by false alarm rate $P_{fa}$ and detection probability $P_d$. For sequential detection (the number of observations is not fixed), the performance is measured by average detection delay and false alarm rate.

## III. Stochastic Resonance

In this section, we briefly introduce the basics of SR.

### A. Introduction to Stochastic Resonance

An intuitive illustration of SR is depicted in Fig. 2. Considering a particle moving between two potential wells. If the particle is forced by a weak periodic signal, e.g. sinusoidal signal, and some white noise, it will oscillate between the two wells. The dynamics of the system affected by both sinusoidal signal and white noise are given by[1]:

$$\begin{aligned} \dot{x}(t) &= -V'(x) + A\sin(2\pi f t) + \eta(t) \\ &= ax(t) - bx(t)^3 + A\sin(2\pi f t) + \eta(t), \end{aligned} \tag{2}$$

where $A$ is the amplitude of the sinusoidal signal and $f$ is its frequency. The sinusoidal signal is contaminated by a white additive noise $\eta$ having power $D$. Here $a$ and $b$ are real parameters. Let $V(x)$ denote the symmetric bistable potential. Then there are two stable points $x_m = \pm\sqrt{a/b}$ and an unstable point $x_b = 0$ in the potential well $V(x) = -\frac{a}{2}x^2 + \frac{b}{4}x^4$. By the force of periodic signal, $x(t)$ fluctuates between the stable states. Noise-driven switching between the local stable states occurs at the Kramers rate [12]:

$$R = \frac{a}{\sqrt{2}\pi}\exp\left(-\frac{\Delta V}{D}\right), \tag{3}$$

where $\Delta V = \frac{a^2}{4b}$ is the height of the potential barrier between the two stable states ($x_m$, $x_b$ and $\Delta V$ are all shown in Fig. 2). When a proper noise power is chosen, the noise-driven switching between the potential wells can become synchronized with the weak periodic signal, which will be amplified by the noise.

[1]There are many possible system dynamics. We consider the simplest one in this paper.

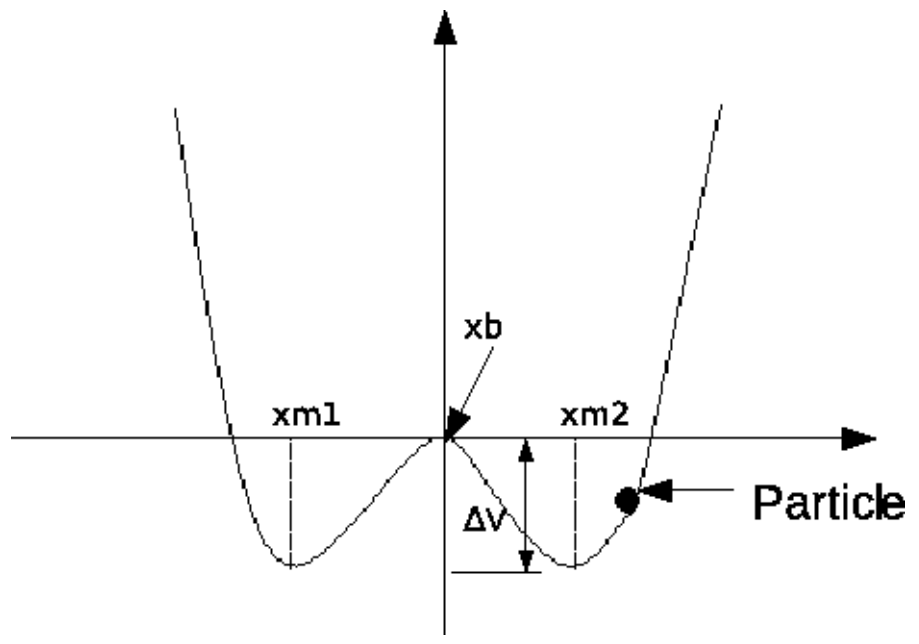

Fig. 2. Illustration of SR in Double Potential Wells

### B. Realization of Stochastic Resonance

There are at least two main approaches to achieve the gain of SR [13] [14] [15]. One approach is fixing the parameters of the SR system and adjusting the input noise level. Another method is dynamically tuning the parameters of SR system, whose implementation is much more complicated. Combinations of these two approaches and other adaptive algorithms have been proposed and studied.

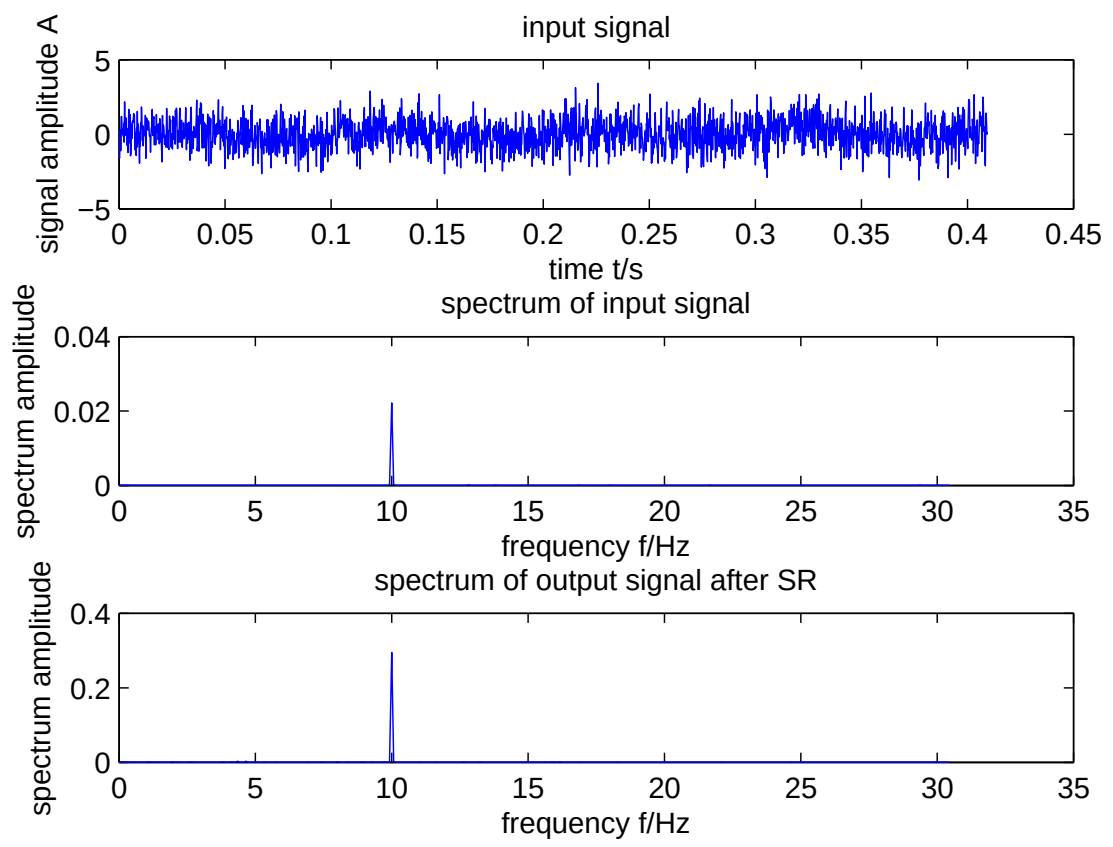

Fig. 3. Comparison of Power Spectral Density

Here we only consider the first approach and apply it in our proposed algorithm. In Eq. (2), on assuming that the signal frequency $f$ is 10Hz and the amplitude $A$ is 0.3, the parameters of the bistable system $V(x)$ are fixed by $a = 1$ and $b = 1$. In order to get the maximum processing gain of SR, we need to find the optimal noise $\eta$. After a simple iterative computation (the details are omitted due to limited space), we can find that the optimal noise power is $D = 0.43$.

Fig. 3 shows the comparison of power spectral density (PSD) between the input signal and output signal processed by SR system. It is obvious that, with the optimal noise power, the spectral power of the output signal at frequency 10Hz increases tremendously (roughly 0.3 while it is 0.02 for the input signal). This is also an illustration that the SR system converts noise power into signal power and thus improves the output SNR.

From Fig. 3, we observe that the SR based nonlinear system can improve SNR when satisfying some specific conditions. This characteristic would be helpful to overcome the $SNR_{wall}$ limitation in energy detection [7] and thus detect weak PU signal.

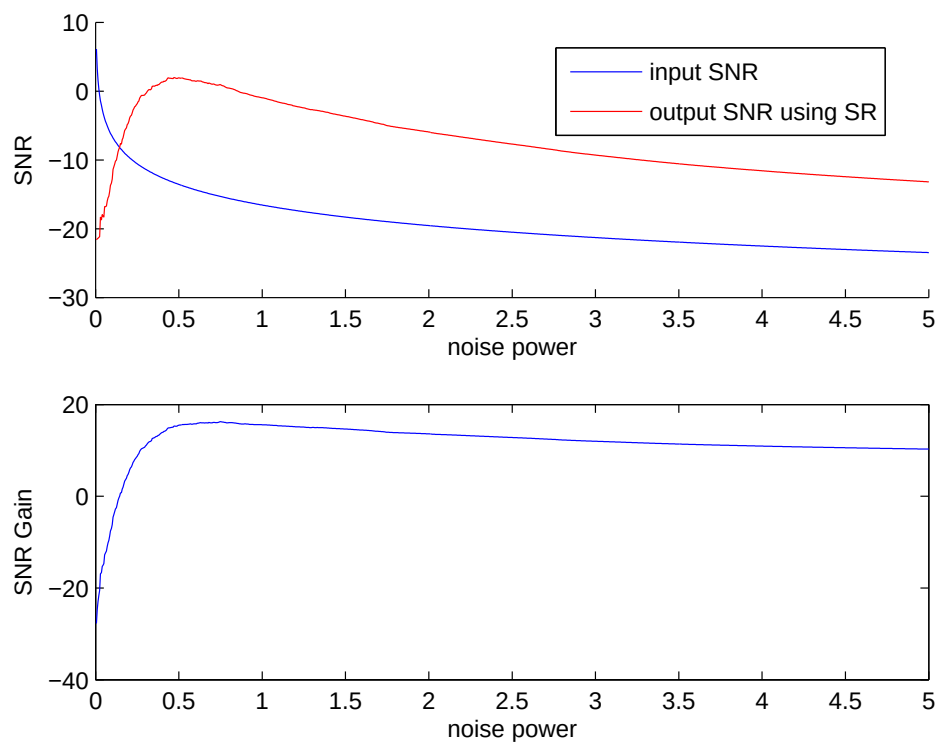

Fig. 4. SNR gain under different Noise levels

Although the SR system has the optimal performance under a certain power level, the SNR is also improved over a wide range of noise levels. Considering the same setting we used to show the PSD of output signal processed by SR system. We can generate input SNR and output SNR under different noise power levels. The simulation result is shown in Fig. 4. We observe that the performance gain is positive for most of the noise levels shown in the figure.

## IV. Stochastic Resonance for Spectrum Sensing

In this section, we first propose two schemes of SR for spectrum sensing in the low SNR regime. Then some implementation issues are discussed.

### A. Block Spectrum Sensing

In order to utilize the processing gain of SR system, we propose an FFT based implementation. The natural idea to use SR before detection is illustrated in Fig. 5. This is motivated by the schemes proposed by Zozor and Inchiosa [16] [17]. Here, we use a specific SR system to pre-treat the input signal and then a simple energy detection [18] is exploited.

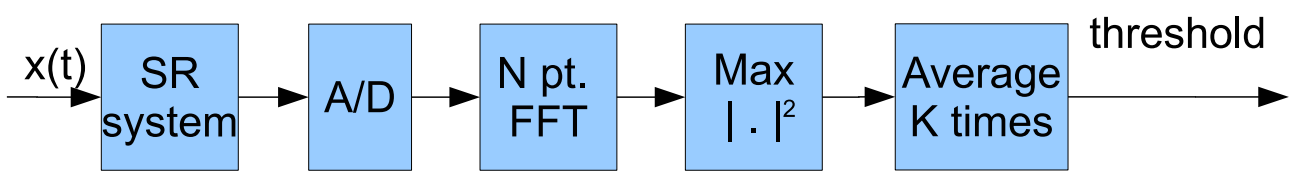

Fig. 5. SR as a pre-treatment of the Energy Detector

For block spectrum sensing in which the decision is made based on the observations within a sensing window, the decision metric for energy detector is:

$$M = \frac{1}{K}\sum_{n=1}^{K}(\max_{t}\{|Y(n,t)|^2\}) \tag{4}$$

where $K$ is the number of FFT blocks within one sensing window and $Y(n,t)$ is the $t$-th output of the $n$-th FFT block.

For each $P_{fa}$ and $P_d$ pair, we can set a corresponding threshold $\gamma$: if $M > \gamma$, we claim that PUs are present. Otherwise, we claim that PUs are absent.

### *B. Sequential Spectrum Sensing*

An alternative approach is sequential detection via stochastic resonance. In this approach, we monitor the output energy of the SR at the PU signal's frequency and set a reference level $E_0$. A heuristic algorithm[2] similar to cumulative sum (CUSUM) test in quickest change detection [19][20] is proposed. On defining $E_n = \max_t |Y(n,t)|^2$ ($Y(n,t)$ is the same as in (4)) for FFT window $n$, the following metric is computed (here we assume that there is no primary user at the beginning):

$$m(n) = \max\{m(n-1) + E_n - E_0, 0\}.$$

When $m(n) > \gamma$, where $\gamma$ is a threshold, we claim that a PU has emerged. Obviously, when primary user emerges, the metric increases in a statistical sense; when there is no primary user, the metric returns to 0 with large probability.

### *C. Implementation Issues*

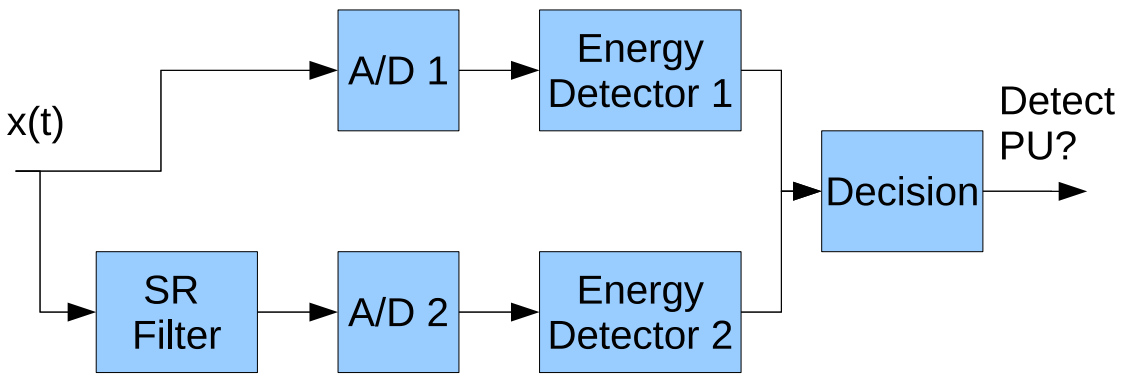

Fig. 6. SR based Energy Detection system

One problem of using SR has been mentioned before, i.e. a specific SR system may not work in the high SNR regime. A feasible approach is depicted in Fig. 6. Two energy detectors, i.e. SR based and ordinary energy detectors, are used simultaneously. The final decision should be a combination of the outputs of the two energy detectors.

Another challenge for applying SR in spectrum sensing is that SR system is usually suitable for weak signal detection with a low frequency. Though it is possible to apply SR in high frequency signal detection, the hardware implementation is more challenging than in low frequency situations. In practice, take ATSC as an example, there are two possible pilots that can be used for sensing obligation, at frequencies 309440.6Hz and 328843.6Hz, respectively. The error of the frequeuncies is within 10Hz. Therefore, we can use a mixer having fixed frequency and a low pass filter to transform the pilot from high frequency to low frequency, e.g. within 10Hz, such that SR can be used to amplify the PU signal.

[2]We are unable to obtain the distribution of the output signal of SR system. Therefore, we cannot apply exact CUSUM test. However, we can replace the likelihood used in CUSUM test with the difference between sensed energy and a reference energy.

## V. Numerical Results

Numerical results are used to test the performance of the spectrum sensing algorithm proposed in this paper. In the simulation, we hold the assumption that the PU signal is a sine signal with frequency 10Hz (transformed from radio frequency) and SR filter is configured with the same setting in the previous section.

### *A. Block Detection*

In Fig. 7, the performance of ordinary energy detector and SR pre-treated energy detector are compared when the size of FFT window is 256 and sensing window is 512.

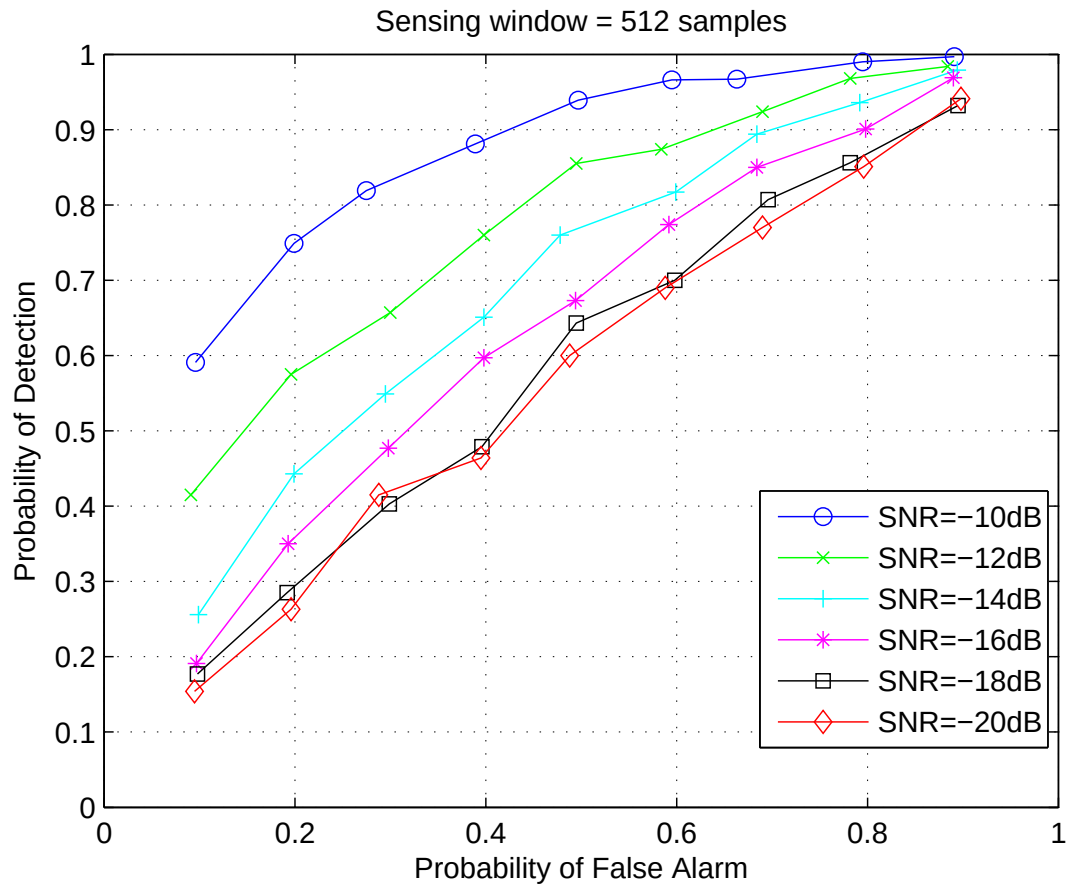

(a) Energy detection

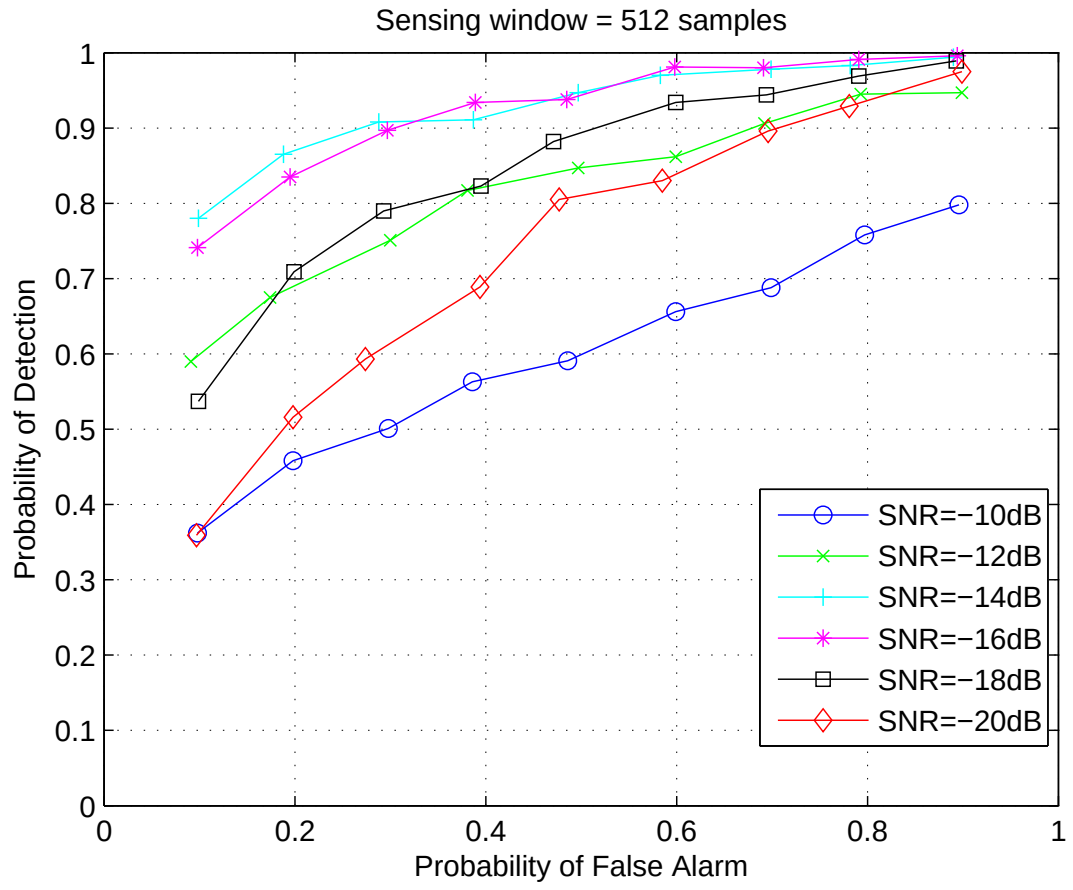

(b) SR pre-treated energy detection

Fig. 7. Comparison of ROC curves between ordingary energy detector and SR based detector

In order to accurately estimate $P_d$ and $P_{fa}$, we repeat each detection measurement for 1000 times. It is obvious that, in the low SNR regime, SR pre-treated energy detector performs much better than the original energy detector. Note that when $SNR$ is not quite low, e.g. $-10dB$, the SR pre-treated approach has a worse performance than the original energy detector.

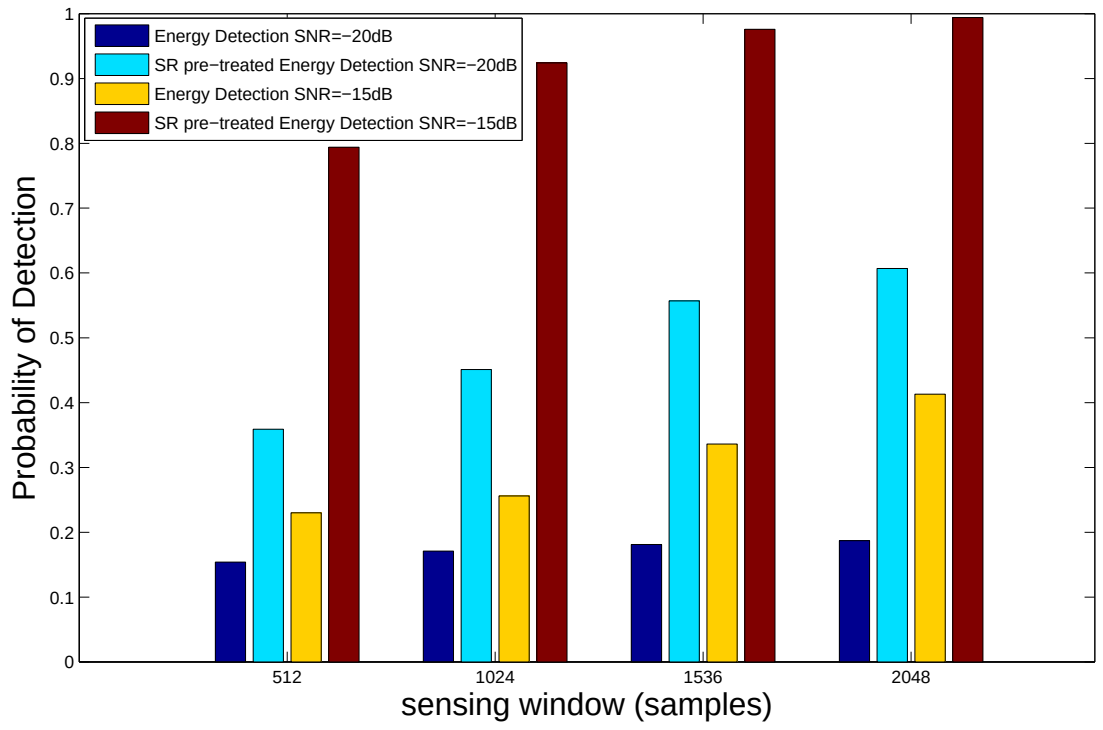

Fig. 8. Detection rate $P_d$ vs sensing time

When $P_{fa}$ is fixed at $10\%$, $P_d$ is obtained using various sensing windows under a low $SNR = -20dB$ which is plotted in Fig. 8. We observe that increasing the number of sensing samples does not significantly improve the detection in the original energy detection when the SNR is very small. In contrast to the original energy detection, $P_d$ obtained by the SR pre-treated energy detector increases much faster when longer sensing time is used. This is because a properly designed SR system can largely improve SNR for the energy detector.

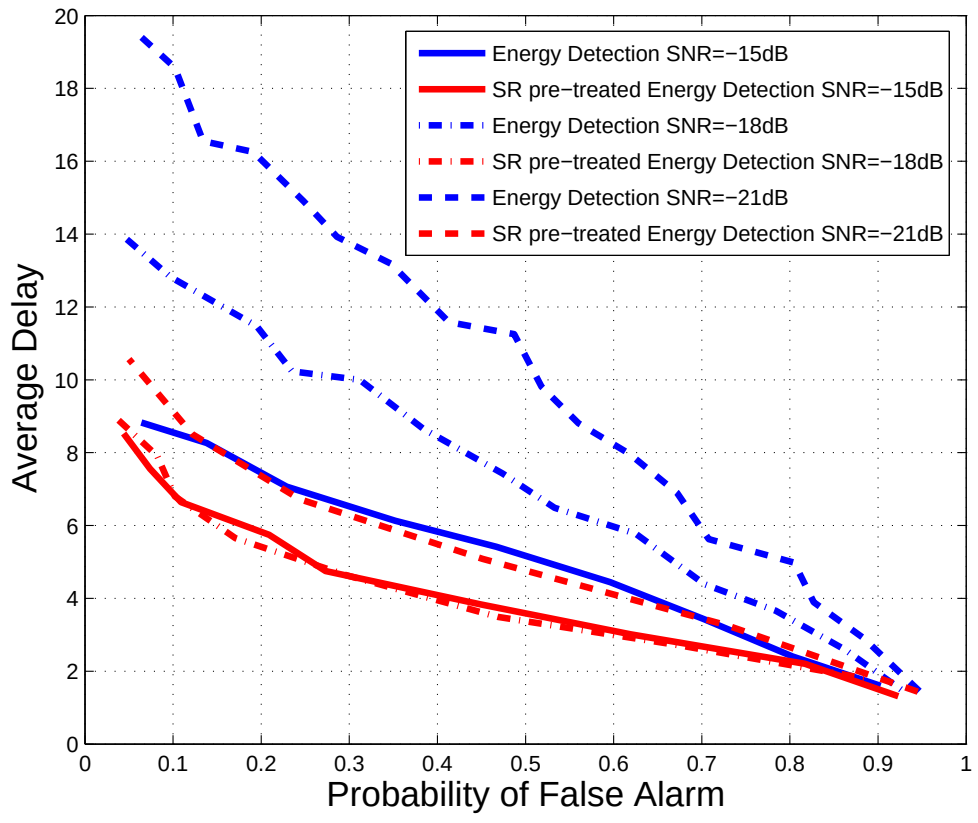

Fig. 9. False alarm rate $P_{fa}$ vs detection delay

### *B. Sequential Detection*

Comparison betwen our proposed SR based sequential detection algorithm with pure energy detection is showed in Fig. 9. The average delay is measured by the number of FFT window used to detect PUs signal. In the low SNR regime, under the same $P_{fa}$, SR based sequential detection will need much less time (averagely) to detect the PUs signal.

## VI. CONCLUSION

In this paper, we have briefly reviewed several common spectrum sensing techniques and pointed out their essential problems. To combat the challenging problem of detecting PUs in the low SNR regime, we have proposed to use SR to enhance the signal SNR under the assumption of periodic PU signal. Both block and sequential detection schemes, based on SR, are proposed. Simulation results show that the proper use of SR pre-treated spectrum sensing can achieve significantly better performance when the PU signal is very weak. Our future work will focus on applying various SR systems in different PUs signal detection schemes and making it more applicable for spectrum sensing.

## REFERENCES

[1] J. Mitola III, "An integrated agent architecture for software defined radio," Ph.D. dissertation, Royal Institute of Technology (KTH), Stockholm, Sweden, May 2000.
[2] T. Yucek and H. Arslan, "A survey of spectrum sensing algorithms for cognitive radio applications," *Communications Surveys & Tutorials, IEEE*, vol. 11, no. 1, pp. 116–130, 2009.
[3] C. R. Stevenson, C. Cordeiro, E. Sofer, and G. Chouinard, "Functional requirements for the 802.22 wran standard," IEEE 802.22-05/0007r47, Jan 2006.
[4] D. Cabric, A. Tkachenko, and R. Brodersen, "Spectrum sensing measurements of pilot, energy, and collaborative detection," in *Military Communications Conference, 2006. MILCOM 2006. IEEE*, Oct. 2006, pp. 1–7.
[5] D. Cabric, S. M. Mishra, and R. W. Brodersen, "Implementation issues in spectrum sensing for cognitive radios," in *Signals, Systems and Computers, 2004. Conference Record of the Thirty-Eighth Asilomar Conference on*, vol. 1, 2004, pp. 772–776 Vol.1.
[6] S. J. Shellhammer, S. S. N, R. Tandra, and J. Tomcik, "Performance of power detector sensors of dtv signals in ieee 802.22 wrans," in *TAPAS '06: Proceedings of the first international workshop on Technology and policy for accessing spectrum*. New York, NY, USA: ACM, 2006, p. 4.
[7] A. Sahai, N. Hoven, and R. Tandra, "Some fundamental limits on cognitive radio," in *Proc. of Allerton Conference*, Oct 2004.
[8] S. Shellhammer and R. Tandra, "An evaluation of dtv pilot power detection," IEEE 802.22-06/0188r0, July 2006.
[9] T. Yucek and H. Arslan, "Spectrum characterization for opportunistic cognitive radio systems," in *Military Communications Conference, 2006. MILCOM 2006. IEEE*, Oct. 2006, pp. 1–6.
[10] J. G. Proakis, *Digital Communications*, 4th ed. McGraw-Hill, 2001.
[11] R. Benzi, A. Sutera, and A. Vulpiani, "The mechanism of stochastic resonance," *J. Phys. A: Math. General*, vol. 14, pp. 453–457, 1981.
[12] H. Risken, *The Fokker-Planck equation*. Springer-Verlag, 1984.
[13] Q. Ye, H. Huang, X. He, and C. Zhang, "A study on the parameters of bistable stochastic resonance systems and adaptive stochastic resonance," in *Robotics, Intelligent Systems and Signal Processing, 2003. Proceedings. 2003 IEEE International Conference on*, vol. 1, Oct. 2003, pp. 484–488 vol.1.
[14] W. Zhao, L. Guo, F. Tian, and L. Shao, "The adaptive detection and application of weak signal based on stochastic resonance," in *Mechatronic and Embedded Systems and Applications, Proceedings of the 2nd IEEE/ASME International Conference on*, Aug. 2006, pp. 1–3.
[15] Z. Hou, J. Yang, Y. Wang, and K. Wang, "Weak signal detection based on stochastic resonance combining with genetic algorithm," in *Communication Systems, 2008. ICCS 2008. 11th IEEE Singapore International Conference on*, Nov. 2008, pp. 484–488.
[16] S. Zozor and P.-O. Amblard, "On the use of stochastic resonance in sine detection," *Signal Process.*, vol. 82, no. 3, pp. 353–367, 2002.
[17] M. E. Inchiosa and A. R. Bulsara, "Signal detection statistics of stochastic resonators," *Phys. Rev. E*, vol. 53, no. 3, pp. R2021–R2024, Mar 1996.
[18] C. Cordeiro, M. Ghosh, D. Cavalcanti, and K. Challapali, "Spectrum sensing for dynamic spectrum access of tv bands," in *Cognitive Radio Oriented Wireless Networks and Communications, 2007. CrownCom 2007. 2nd International Conference on*, Aug. 2007, pp. 225–233.
[19] H. V. Poor and O. Hadjiliadis, *Quickest Detection*. Cambridge University Press, 2008.
[20] L. Lai, Y. Fan, and H. V. Poor, "Quickest detection in cognitive radio: A sequential change detection framework," in *Proc. of IEEE Global Communication Conference*, 2008.